\documentclass[a4paper, 10pt, conference]{ieeeconf}

\IEEEoverridecommandlockouts % This command is only needed if you want to use the \thanks command
\usepackage{graphics} % for pdf, bitmapped graphics files
\usepackage{epsfig} % for postscript graphics files
\title{\LARGE \bf
Optimal filtering techniques for the adaptive optics system of the LBT
}

\author{G. Agapito, F. Quir\'os-Pacheco, P. Tesi, A. Riccardi, S. Esposito% <-this % stops a space
\thanks{G. Agapito, F. Quir\'os-Pacheco, A. Riccardi, S. Esposito are with Osservatorio Astrofisico di Arcetri, Largo E. Fermi 5, Firenze, Italy}%
\thanks{P. Tesi is with Universit\`a degli Studi di Firenze, Facolt\`a di Ingegneria, via Santa Marta 3, Firenze, Italy}%
\thanks{G. Agapito: E-mail:{\tt\small agapito@arcetri.astro.it}}
\thanks{F. Quir\'os-Pacheco: E-mail:{\tt\small fquiros@arcetri.astro.it}}        
}

\begin{document}

\maketitle
\thispagestyle{empty}
\pagestyle{empty}

\begin{abstract}

In this paper we will discuss the application of optimal filtering techniques for the adaptive optics system of the LBT telescope. We have studied the application of both Kalman and $H_{\infty}$ filters to estimate the temporal evolution of the phase perturbations due to the atmospheric turbulence and the telescope vibrations on tip/tilt modes. We will focus on the $H_{\infty}$ filter and on its advantages and disadvantages over the Kalman filter.

\end{abstract}

\section{INTRODUCTION}

The Large Binocular Telescope (LBT) is an optical/infrared telescope using two 8.4m diameter primary mirrors.  By having both primary mirrors on the same mechanical mount, LBT will be able to achieve the diffraction-limited image sharpness of a 22.8m diameter aperture. As in any large ground-based telescope, the diffraction limit can only be obtained with the assistance of adaptive optics (AO), which is a technique aimed at reducing the effects of wavefront distortion due to atmospheric turbulence~\cite{RODDIER1999}.

\begin{figure}
\begin{center}
\includegraphics[width=8cm]{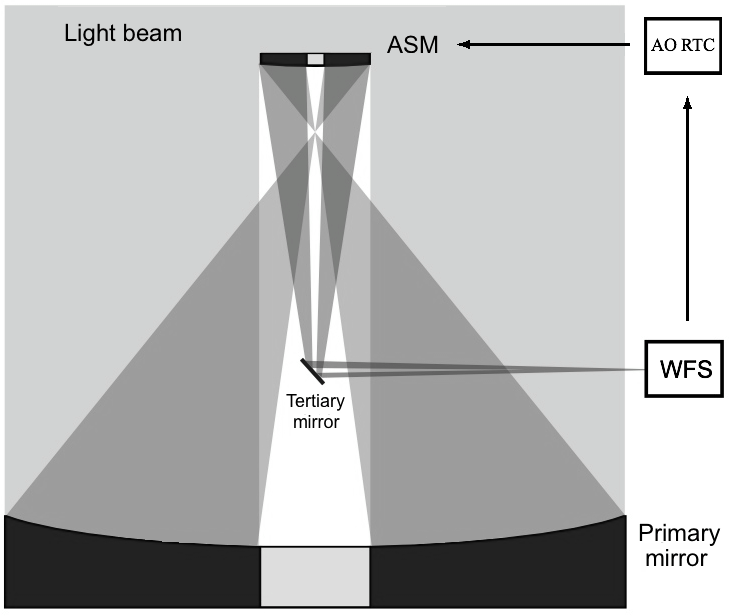}
\caption{Illustration of the optical configuration of one-arm of the LBT, including the AO system components: the wavefront sensor (WFS), the adaptive secondary mirror (ASM), and the real-time computer (RTC).}
\label{wfs}
\end{center}
\end{figure}

LBT will be equipped soon\footnote{at the beginning of 2009 the first LBT-AO system will be commissioned to the telescope.}  with two AO systems, one for each arm of the telescope. Each AO unit (fig.\ref{wfs}) comprises a pyramid wavefront sensor (WFS), an adaptive secondary mirror (ASM), and a real-time computer (RTC). The pyramid wavefront sensor delivers a signal that is proportional, as a first-order approximation, to the first derivative of the incoming wavefront, sampled with a maximum of $30 \times 30$ subapertures~\cite{pyr}.
%is made of a double pyramid prism with a $10\mu m$ vertex manufacturing. The first prism has an angle of $30^{\circ}$, the %second one has an angle of $28^{\circ}$. 
The ASM is a deformable mirror with 672 voice-coil (electro-magnetic force) actuators, distributed in concentric rings, to change the shape of the 1.6mm-thick and 911mm-diameter Zerodur shell~\cite{acc_test}. 
%Finally, the RTC is a DSP-based custom system

Large telescopes suffer from structure vibrations that can reduce the AO performance~\cite{naos}. Recent theoretical studies and preliminary laboratory validations have shown that optimal control techniques can be used to reduce the impact of these vibrations~\cite{validation,kal}. We will present in this paper an analysis of a mixed-control strategy for the LBT based on both optimal filtering and classical control techniques, aimed at reducing the impact of telescope vibrations without burdening the RTC with heavy computations. In section~\ref{strategy} we will present the general control strategy for the LBT-AO system. Section~\ref{model} describes the models required to design the filter-based controllers. We have compared the performance of these controllers based on numerical simulations. These results will be presented in section~\ref{simulation}.

\section{GENERAL CONTROL STRATEGY}\label{strategy}

\begin{figure*}[!t]
%\begin{figure}
\begin{center}
\includegraphics[width=13cm]{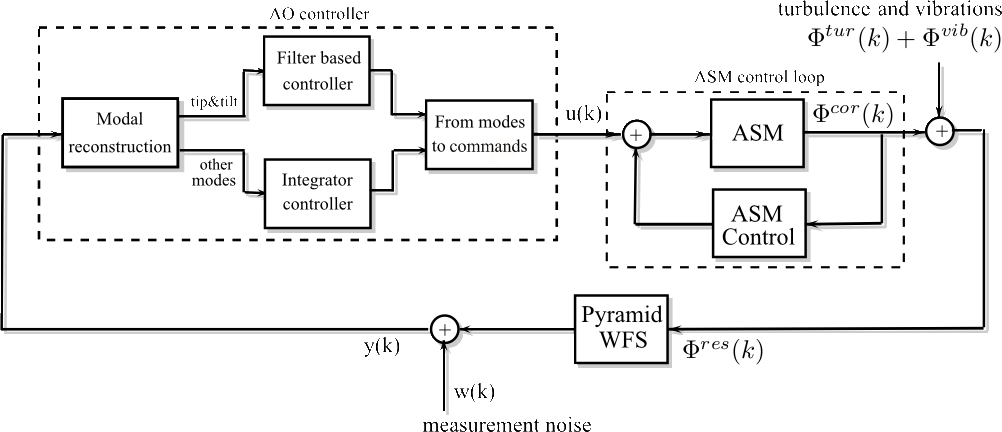}
\caption{LBT-AO system control loop scheme.}
\label{control}
\end{center}
\end{figure*}
%\begin{figure}
%\begin{center}
%\includegraphics[width=9cm]{schema_controllo_attuatore.png}
%\caption{ASM control loop scheme.}
%\label{ASMcontrol}
%\end{center}
%\end{figure}

The AO control diagram for the LBT is illustrated in figure~\ref{control}. The AO controller receives the WFS measurements $y(k)$ and computes the commands vector $u(k)$ to drive the actuators of the ASM. It is important to mention that the ASM has non-negligible dynamics and, to compensate for wavefront distortions, it must take the desired shape with good accuracy and within a short settling time. For this reason, it was chosen to control the ASM with a dedicated control loop. The ASM control loop relies on the position feedback provided by a set of capacitive sensors placed at the back of the mirror shell. The design of the ASM controller is based on a proportional-derivative (PD) position feedback plus a feedforward signal that is proportional to the desired position~\cite{riccardi2003}. Summing up, there are two control systems involved: 
\begin{itemize}
\item A global AO control system (working $@1$kHz) whose goal is to determine the commands to the ASM that corrects for a residual wavefront distortion;
\item A local ASM control system (working $@72$kHz) with the goal of shaping the mirror in the time of one AO loop step ($<1ms$).
\end{itemize}%

The main subject of this paper regards the controller for the global AO loop. We will follow the \emph{modal control approach} widely used in the analysis of AO systems~\cite{RODDIER1999}. The controllers implemented in the current generation of AO systems are based on classical (modal) integrators~\cite{GENDRON1994}. These controllers have provided good performance on atmospheric turbulence correction, but they have been unable to attenuate substantially the effects of telescope structure vibrations. In the LBT case, the swinging arm supporting the ASM has resonance frequencies in the band between $15$ and $30Hz$~\cite{tec}. These vibrations affect mostly the tip/tilt modes\footnote{Tip and tilt modal coefficients quantify the displacements of the image in the two orthogonal directions.}. As we will discuss in section~\ref{simulation}, the vibration attenuation performed by classical controllers is not enough to meet the expected AO performance. For this reason it has been chosen to control tip/tilt modes with a filter-based control. We will review in section~\ref{model} the models of the AO system and of the input signals (atmospheric turbulence, telescope vibrations, measurement noise, etc.) required to define a system state vector and estimate its evolution with an optimal filtering technique such as Kalman or $H_{\infty}$ one (see Agapito~\emph{et.~al.}~2008~\cite{kalhinf} for further considerations).

Finally, let us emphasize that the AO controller studied in this work is a mixed controller (see fig.~\ref{control}): tip/tilt modes are controlled by a Kalman or $H_{\infty}$ filter-based controller whereas the other modes are controlled with a simple integrator controller. The modal basis we chose was created from Karhunen-Lo\`{e}ve modes~\cite{RODDIER1999} defined in the LBT pupil, projected onto the ASM influence functions and then re-orthonormalized. A total of 672 modes (corresponding to the total number of ASM actuators and hence, the total number of degrees of freedom) were computed in this way. Finally, tip/tilt modes were projected out from all modes in order to decouple the control of tip/tilt and the rest of the modes for the mixed-control strategy implementation.

\section{AO SYSTEM MODEL} \label{model}

\subsection{WFS and ASM models}
The pyramid WFS model is described by:
\begin{equation} \label{eq:ao5}
y(k)=D\Phi^{res}(k-1)+w(k)
\end{equation}
where $y(k)$ is the measurement vector, $w(k)$ is the measurement noise vector ---$y(k)$ and $w(k)\in\mathbf{R}^{q \times 1}$ where $q$ is the number of measurements---, $D$ is the WFS response matrix, and $\Phi^{res}(k)$ stands for the residual phase after ASM correction computed as $\Phi^{res}(k)=\Phi^{tot}(k)-\Phi^{cor}(k)$, where $\Phi^{cor}(k)$ is the phase correction applied by the ASM and $\Phi^{tot}(k) = \Phi^{tur}(k) + \Phi^{vib}(k)$, i.e.~the sum of the phase distortions introduced by the turbulence $\Phi^{tur}(k)$ and the telescope vibrations $\Phi^{vib}(k)$. All phase variables are modal coefficient vectors $\Phi(k)\in\mathbf{R}^{n \times 1}$ where $n$ is the number of coefficients.

The ASM model can be expressed by:
\begin{equation} \label{eq:ao1}
\Phi^{cor}(k-1)=Nu(k-2)
\end{equation}
where $N$ is the ASM influence matrix, and $u(k)$ is the command vector for the actuators of the ASM ---$u(k)\in\mathbf{R}^{m \times 1}$, where $m$ is the number of actuators. Note that this equation does not take into account for the mirror dynamics. However, the AO command vector $u(k)$ becomes the reference to the ASM control loop and, as we mentioned above, this loop guarantees that the ASM takes the desired shape. 

\subsection{Turbulence and vibration models}
The atmospheric turbulence evolution can be described by:
\begin{equation} \label{eq:ao2}
\Phi^{tur}(k+1)=f(\Phi^{tur}(k),\Phi^{tur}(k-1),\ldots)+v(k)
\end{equation}
where $v(k)$ is the model's white noise. We have chosen to approximate this equation with an Auto-Regressive (AR) first-order model~\cite{sys}:
\begin{equation} \label{eq:ao3}
\Phi^{tur}(k+1)=A_t\Phi^{tur}(k)+v_t(k)
\end{equation}
where $A_t$ is a diagonal matrix calculated as in Le~Roux~\emph{et.~al.}~2004~\cite{opt}, whose diagonal elements are $e^{-2 \pi 0.3 \eta V/f}$ ($\eta$ radial order, $V$ wind speed, $f$ sampling frequency), and $v_t(k)$ is the model's white noise calculated from the Noll matrix~\cite{noll}.

The vibrations model can be expressed as:
\begin{equation} \label{eq:ao4}
\Phi^{vib}(k+1)=A_{1}\Phi^{vib}(k)-A_{2}\Phi^{vib}(k-1)+v_v(k)
\end{equation}
where $A_{1}$ and $A_{2}$ are two diagonal matrices whose diagonal elements depend upon vibration frequency and damping constant, and $v_v(k)$ is a white noise vector whose variance depends upon input force power~\cite{validation}. $\Phi^{vib}(k)$ and $v_v(k)$ have $p$ non-zero elements corresponding to the modes affected by vibrations.

\subsection{Classical control strategy} \label{integrator}

\begin{figure}
\begin{center}
\includegraphics[width=8cm]{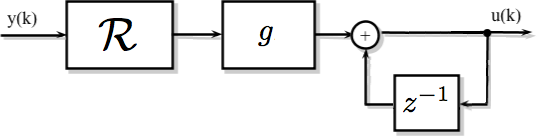}
\caption{Classical control strategy scheme.}
\label{Icontrol}
\end{center}
\end{figure}

The classical control strategy (see figure~\ref{Icontrol}) is based on a reconstruction matrix $\mathcal{R}$ and on a simple integrator:
\begin{equation} \label{eq:sys9}
u(k)=u(k-1)+g\Delta u(k-2)
\end{equation}
where $g$ is the integrator gain (equal for all modes), and the command increment is computed as:
\begin{equation} \label{eq:sys8}
\Delta u(k)=\mathcal{R}y(k) \, .
\end{equation}
The reconstruction matrix $\mathcal{R}$ is computed as $\mathcal{R}=(M'_{int}M_{int})^{-1}M'_{int}$, that is, the generalized inverse of the interaction matrix $M_{int}=DN$, measured experimentally during the AO system calibration. This is the control applied to all modes except for tip/tilt in the mixed-control strategy. We should note that the gain $g$ can be optimized for each mode, as in the case of the optimized modal gain integrator (OMGI)~\cite{GENDRON1994} controller. In this work we did not implement it, but this will be considered as a future improvement.

\subsection{Filter-based control strategy} \label{filtercontrol}

\begin{figure}
\begin{center}
\includegraphics[width=8cm]{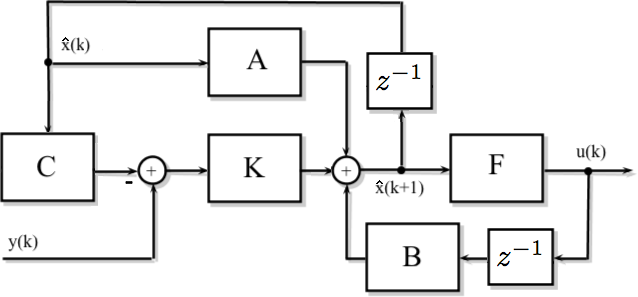}
\caption{Filter-based control strategy scheme (K is the filter asymptotic gain matrix).}
\label{Fcontrol}
\end{center}
\end{figure}

The control based on the Kalman or the $H_{\infty}$ filter (see figure \ref{Fcontrol}) generates the command vector from the predicted state vector. We have defined the following state vector:
\begin{equation}\label{eq:sys2}
x(k)=\left[ \begin{array}{c}\Phi^{vib}(k)\\ \Phi^{vib}(k-1)\\ \Phi^{tur}(k)\\ \Phi^{tot}(k-1)\\ u(k-2) \end{array} \right]
\end{equation}
comprising all the variables required to estimate the total phase vector $\Phi^{tot}(k+1)$. The dimension of the state vector is $(2p+2n+m) \times 1$. It turns out that the command vector $u(k)$ is computed by projecting $\hat{\Phi}^{tot}(k+1)$ onto the command space:
\begin{equation} \label{eq:sys6}
\begin{array}{ll}
u(k) & = (N'N)^{-1}N' F \hat{x}(k+1)\\
 & =(N'N)^{-1}N'\hat{\Phi}^{tot}(k+1) \, ,
\end{array}
\end{equation}
where $F=\left[ \begin{array}{ccccc} I & 0 & I & 0 & 0 \end{array} \right]$ and $I$ is the identity matrix.

For the Kalman filter, the state model is expressed as:
\begin{equation}\label{eq:sys1}
\begin{array}{ll}
x(k+1)&=\underbrace{\left[ \begin{array}{ccccc}A_{1} & A_{2} & 0 & 0 & 0\\ I & 0 & 0 & 0 & 0\\ 0 & 0 & A_t & 0& 0\\I & 0 & I & 0 & 0\\0 & 0 & 0 & 0 & 0 \end{array} \right]}_{A}x(k)+\\
&\quad \underbrace{\left[ \begin{array}{c} 0\\ 0\\ 0\\ 0\\ I \end{array} \right]}_{B}u(k-1)+v(k) \, ,
\end{array}
\end{equation}
where $v(k)=\left[ \begin{array}{ccccc} v_v(k) & 0 & v_t(k) & 0 & 0 \end{array} \right]$. Finally, the measurement equation is expressed as:
\begin{equation}\label{eq:sys1a}
\begin{array}{ll}
y(k)&=\underbrace{D \left[ \begin{array}{ccccc} 0 & 0 & 0 & I & -N \end{array} \right]}_{C}x(k)+w(k) \, .
\end{array}
\end{equation}

On the other hand, the state model for the $H_{\infty}$ filter is expressed as:
%\begin{figure*}[!t]
%% ensure that we have normalsize text
%\normalsize
%% Store the current equation number.
%\setcounter{MYtempeqncnt}{\value{equation}}
%% Set the equation number to one less than the one
%% desired for the first equation here.
%% The value here will have to changed if equations
%% are added or removed prior to the place these
%% equations are referenced in the main text.
%\setcounter{equation}{1}
%\begin{equation}
%\left[ \begin{array}{c} x(k+1)\\ \hat{z}(k)-z(k) \\ y(k) \end{array} \right]=
%\left[ \begin{array}{ccccc|ccc|c}A_{1} & A_{2} & 0 & 0 & 0 & I & 0 & 0 & 0\\ I & 0 & 0 & 0 & 0 & 0 & 0 & 0 & 0\\ 0 & 0 & A_t & 0 & 0 & 0 & I & 0 & 0\\I & 0 & I & 0 & 0 & 0 & 0 & 0 & 0\\0 & 0 & 0 & 0 & 0 & 0 & 0 & 0 & 0\\ \hline -I & 0 & -I & 0 & 0 & 0 & 0 & 0 & I\\ \hline 0 & 0 & 0 & D & -DN & 0 & 0 & I & 0\\ \end{array} \right]\left[ \begin{array}{c} x(k)\\ \mu(k) \\ \hat{z}(k) \end{array} \right]+
%\left[ \begin{array}{c} 0\\ 0\\ 0\\ 0\\ I\\ \hline 0\\ \hline 0 \end{array} \right]u(k-1)  \, ,
%\end{equation}
%% Restore the current equation number.
%\setcounter{equation}{\value{MYtempeqncnt}}
%% IEEE uses as a separator
%\hrulefill
%% The spacer can be tweaked to stop underfull vboxes.
%\vspace*{4pt}
%\end{figure*}
\begin{equation}\label{eq:sys3}
\begin{array}{l}
\left[ \begin{array}{c} x(k+1)\\ \hat{z}(k)-z(k) \\ y(k) \end{array} \right]=\\ \\
\left[ \begin{array}{ccccc|ccc|c}A_{1} & A_{2} & 0 & 0 & 0 & I & 0 & 0 & 0\\ I & 0 & 0 & 0 & 0 & 0 & 0 & 0 & 0\\ 0 & 0 & A_t & 0 & 0 & 0 & I & 0 & 0\\I & 0 & I & 0 & 0 & 0 & 0 & 0 & 0\\0 & 0 & 0 & 0 & 0 & 0 & 0 & 0 & 0\\ \hline -I & 0 & -I & 0 & 0 & 0 & 0 & 0 & I\\ \hline 0 & 0 & 0 & D & -DN & 0 & 0 & I & 0\\ \end{array} \right]\left[ \begin{array}{c} x(k)\\ \mu(k) \\ \hat{z}(k) \end{array} \right]\\ \\
+ \left[ \begin{array}{c} 0\\ 0\\ 0\\ 0\\ I\\ \hline 0\\ \hline 0 \end{array} \right]u(k-1)  \, ,
\end{array}
\end{equation}
with $x(0) = 0$. Also, $z(k)= \Phi^{tot}(k+1)$, $\hat{z}(k)$ is an estimate of $z(k)$, and $\mu(k)$ is the disturbances vector $\mu(k)=\left[ \begin{array}{ccc} v_v(k) & v_t(k) & w(k) \end{array} \right]'$.

\begin{table*}
\begin{center}
\begin{tabular}{cc}
\begin{tabular}{|l|c|}
\hline
\multicolumn{2}{|c|} {\textbf{Telescope}} \\
\hline
Effective diameter ($D$) & $8.22m$ \\
\hline
Central obstruction & $0.11 D$ \\
\hline
\multicolumn{2}{|c|}{\textbf{Pyramid WFS}}\\
\hline
Sensing wavelength ($\lambda$) & $0.75\mu m$\\
\hline
Tilt modulation radius & $4.0\frac{\lambda}{D}$\\
\hline
Number of subapertures & $30 \times 30$\\
\hline
Number of photons per integration time per subaperture & $50$\\
\hline
Number of electrons per pixel of readout noise & $8$ \\
\hline
\end{tabular} & \begin{tabular}{|l|c|}
\hline
\multicolumn{2}{|c|} {\textbf{ASM}}\\
\hline
Number of modes & $672$ \\
\hline
\multicolumn{2}{|c|}{\textbf{Turbulence}}\\
\hline
Seeing & $0.8$ ($@ \, 0.5\mu m$)\\
\hline
Outer scale ($L_0$) & $22m$\\
\hline
Wind speed & $20m/s$\\
\hline
\multicolumn{2}{|c|}{\textbf{Loop parameters}}\\
\hline
Sampling frequency & $800Hz$\\
\hline
Total delay & 2 frames\\
\hline
\end{tabular}\\
\end{tabular}
\caption{Summary table of simulation parameters.}
\label{simparams}
\end{center}
\end{table*}

\section{FILTERS}\label{filter}

We decided to implement the Kalman filter because it is the best linear state estimator and the $H_{\infty}$ filter because it is capable of dealing with plant errors and unknown disturbances.
Kalman and $H_{\infty}$ filters have different objectives:
\begin{itemize}
 \item the Kalman filter's aim is to minimize either the variance of the final state estimation error:
\begin{equation} \label{eq:f1}
J_1=\varepsilon\left[(\hat{x}(\texttt{N})-x(\texttt{N}))'(\hat{x}(\texttt{N})-x(\texttt{N}))\right],
\end{equation}
or to minimize the average RMS power of the estimation error~\cite{LRC}:
\begin{equation} \label{eq:f2}
J_2=\varepsilon\left[\frac{1}{\texttt{N}}\sum_{k=0}^\texttt{N}\left(\hat{x}(k)-x(k)\right)'\left(\hat{x}(k)-x(k)\right)\right]^{\frac{1}{2}}
\end{equation}
where $\varepsilon[\cdot]$ denotes the expected value, and $x(\texttt{N})$ denotes the final state;
\item the $H_{\infty}$ filter's aim is to ensure that the energy gain from the disturbances to the estimation error is less than a prespecified level $\gamma^2$~\cite{LRC}:
\begin{equation} \label{eq:f3}
\|\hat{z}-z\|^2_{2,[0,\texttt{N}]}  - \gamma^2\| \mu\|^2_{2,[0,\texttt{N}]}\leq-\epsilon\| \mu\|^2_{2,[0,\texttt{N}]}
\end{equation}
where $\epsilon>0$, $\mu \in l_2[0,\texttt{N}]$ is the disturbances vector\footnote{The space $l_2[0,\texttt{N}]$ is defined as:
\begin{displaymath}
l_2[0,\texttt{N}]=\left\{ f:f(k)=0 \; \forall \; k \notin [0,\texttt{N}], \| f\|_{2,[0,\texttt{N}]}<\infty \right\} \, ,
\end{displaymath} 
where $\|\cdot\|_{2,[0,\texttt{N}]}$ is the finite-horizon 2-norm, defined as:
\begin{displaymath}
\| f\|_{2,[0,\texttt{N}]} = \left\{ \sum_{k=0}^\texttt{N} f'(k)f(k) \right\}^{\frac{1}{2}} \, ,
\end{displaymath}
where $f=\left\{f(k)\right\}_{-\infty}^{\infty}$.}, $z=Lx$, and $\hat{z}=Fy$ is a $z$ estimate ($F$ must be casual and linear).
\end{itemize}

Kalman and $H_{\infty}$ filters use different problem descriptions too:
\begin{itemize}
 \item for  Kalman filter the signal generating system is assumed to be a state-space system driven by a white noise process with known statistical properties. The observed output is also corrupted by a white noise process with known statistical properties;
\item for $H_{\infty}$ filter the system has unknown disturbances of finite energy that drive the signal generating system and corrupt the observations.
\end{itemize}

\section{SIMULATIONS AND DATA ANALYSIS} \label{simulation}

\subsection{Preliminary considerations}
Telescope structure vibrations may exhibit large amplitudes, in particular on tip/tilt modes, and they depend on many factors such as telescope orientation, telescope tracking errors, and wind shaking. 

The LBT relies on a set of accelerometers placed on the structure supporting the ASM to characterize the vibrations (frequency and amplitude) affecting the AO system. In this work we did not consider an adaptive controller, so the vibration's parameters will be previously calibrated with the accelerometers and used to build the Kalman (or $H_{\infty}$) filter. For these reasons it is important to study the robustness of the controllers with respect to errors in the vibration's model.

We will first compare the performance of the three controllers (classical, mixed-Kalman, and mixed-$H_{\infty}$) under the presence of only atmospheric turbulence (sec.~\ref{justturb}). Then, we will consider the presence of a telescope vibration affecting tip-tilt modes (sec.~\ref{turbvib}). Finally, we will study the robustness of the mixed-controllers with respect to changes on the vibration frequency (sec.~\ref{robustness}).

All the simulations were made on an end-to-end simulator of the LBT-AO system. Table~\ref{simparams} presents a summary of the simulation parameters.

\subsection{Performance under the presence of turbulence}\label{justturb}

First, let us consider that there are no vibrations. In this case, mixed-Kalman controller gives a $SR$\footnote{To measure the performance of an AO system we use the Strehl Ratio ($SR$). It is the ratio of the observed peak intensity at the detection plane compared to the theoretical maximum peak intensity of a diffraction-limited image.} of $80.7\%$, the mixed-$H_{\infty}$ controller a $SR$ of $80.4\%$, and the classical controller a $SR$ of $84.1\%$ (Table \ref{Iresults}). Note that the performance of the mixed controllers is slightly lower with respect to the classical one because the AR1 dynamic model of the turbulence is a simple one; it is just a first-order approximation of the Taylor's hypothesis model of the turbulence's temporal evolution~\cite{opt}. 
%Also the mixed controller contain the vibration model that, in this case, brings some errors in the estimation.

\begin{center}
\begin{table}
\begin{center}
\begin{tabular}{|c|c|c|c|}
\hline
{} & \multicolumn{3}{|c|}{\%SR @ $2.2\mu m$}\\
\hline
vibration & Classical & mixed-Kalman & mixed-$H_{\infty}$\\
\hline
No & 84.1 & 80.7 & 80.4\\
\hline
Yes & 30.9 & 80.4 & 80.2\\
\hline
\end{tabular}
\end{center}
\caption{Simulation results: performance of the LBT-AO system with and without telescope vibrations.}
\label{Iresults}
\end{table}
\end{center}

\subsection{Performance under the presence of turbulence and vibrations}\label{turbvib}
Let us now consider the case where there is a telescope vibration affecting tip/tilt modes with an amplitude of $80$~milli\-arcseconds at a frequency of $20Hz$. Under these conditions, mixed-Kalman controller provides a $SR$ of $80.4\%$, and mixed-$H_{\infty}$ controller a $SR$ of $80.2\%$. Their performances are very similar to the ones obtained in the previous case. On the other hand, the classical controller has a very different performance; the $SR$ has been reduced to $30.9\%$ under the presence of this vibration (Table \ref{Iresults}). We should note that this result was obtained by increasing the integrator's gain in order to increase the attenuation at the vibration's frequency. Of course, the gain cannot be increased arbitrarily due to stability constraints. Therefore, the AO performance with the classical controller will remain limited by the presence of telescope vibrations.

\subsection{Robustness study}\label{robustness}
\begin{figure}
\begin{center}
\includegraphics[width=8.5cm]{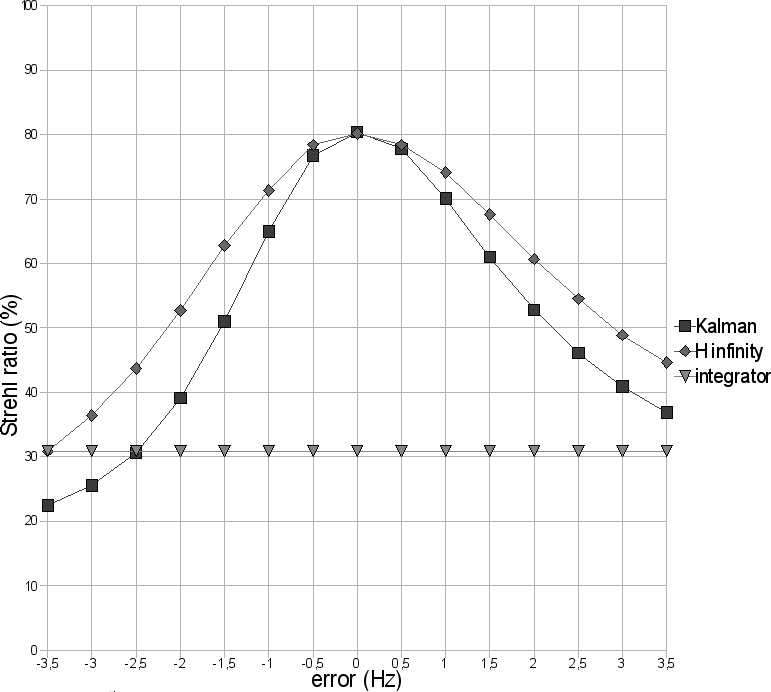}
\caption{Robustness study: performance of the mixed controllers under the presence of model errors regarding the vibration's frequency.}
\label{simul}
\end{center}
\end{figure}

In order to test the robustness of the controllers based on the Kalman and the $H_{\infty}$ filters, we introduced an error on the value of the vibration's frequency in the state model, whereas the actual vibration's frequency was left equal to $20Hz$.
From figure \ref{simul} (and table \ref{IIresults}) we can see that the two filters have very similar performance when the error on the frequency is less than $\left|0.5Hz\right|$. When the error is greater than $\left|1Hz\right|$ the performance of the mixed-$H_{\infty}$ is $\approx 10\%$ in $SR$ better than the mixed-Kalman controller. Note that the $SR$ of the classical controller with this vibration is lower than the mixed controllers almost for every considered error values. 

\begin{table}
\begin{center}
\begin{tabular}{|c|c|c|c|}
\hline
\multicolumn{2}{|c|}{frequency ($Hz$)} &\multicolumn{2}{|c|}{\%SR @ $2.2\mu m$}\\
\hline
model & error & mixed-Kalman & mixed-$H_{\infty}$\\
\hline
16.5 & -3.5 & 22.4 & 30.9\\
17 & -3 & 25.5 & 36.4\\
17.5 & -2.5 & 30.6 & 43.7\\
18 & -2 & 39.1 & 52.7\\
18.5 & -1.5 & 51.0 & 62.8\\
19 & -1 & 65 & 71.3\\
19.5 & -0.5 & 76.8 & 78.4\\
\hline
20 & 0 & 80.4 & 80.2\\
\hline
20.5 & 0.5 & 77.8 & 78.4\\
21 & 1 & 70.1 & 74.1\\
21.5 & 1.5 & 61 & 67.6\\
22 & 2 & 52.8 & 60.7\\
22.5 & 2.5 & 46.1 & 54.5\\
23 & 3 & 40.9 & 48.9\\
23.5 & 3.5 & 36.9 & 44.6\\
\hline
\end{tabular}
\end{center}
\caption{Strehl Ratio values shown in figure~\ref{simul}.}
\label{IIresults}
\end{table}

\begin{figure}
\begin{center}
\includegraphics[width=8.5cm]{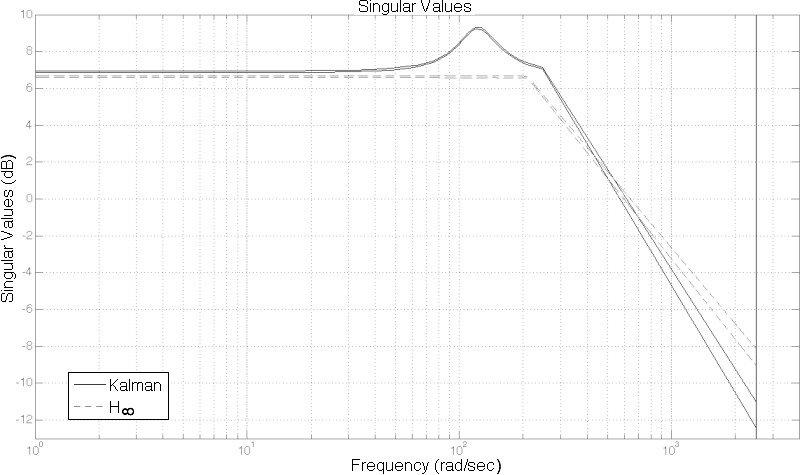}
\caption{Singular values of the \emph{disturbances} - \emph{esitmation error} transfer function for Kalman and $H_{\infty}$ filters.}
\label{sigma}
\end{center}
\end{figure}

These simulation results can also be explained by looking at the corresponding sensitivity functions. Figure \ref{sigma} represents the maximal singular values of the transfer functions between disturbances and estimation error\footnote{We trace this graph and not all the sensitivity functions for a simpler and better comprehension - the sensitivity functions are $n \times m$.}. From this figure we can see that the Kalman filter estimation sensitivity functions in correspondence of the vibration frequency have a peak. This means that the Kalman filter is more sensitive to disturbances around this frequency, and that model errors around this frequency will have a greater influence on the estimation. Instead, the $H_{\infty}$ filter is characterized by flatter sensitivity functions. Hence, this filter should be more robust to errors on the vibration's frequency value, as has been shown with numerical simulations above.

\addtolength{\textheight}{-1.8cm}   % This command serves to balance the column lengths
                                  
\section{CONCLUSIONS AND FURTHER WORK}

We have presented in this work a mixed-control strategy combining classical and filter-based techniques for the LBT-AO system. We have shown with numerical simulations that the mixed controllers are able to effectively eliminate the effects of telescope's structure vibrations on the AO performance. In order to achieve this, it is crucial to characterize accurately the vibration parameters, in particular the vibration's frequency value. We have verified that the $H_{\infty}$ filter is more robust than the Kalman filter with respect to uncertainties on the vibration's frequency value. For the particular parameters simulated in this work, an absolute loss of $10\%$ of $SR$ at 2.2$\mu m$ is expected in the presence of a frequency error of $\pm 1.2Hz$ and $\pm 0.9Hz$ in the vibration's model for the $H_{\infty}$ and the Kalman filter respectively.

We should note that more than one vibration frequencies can be taken into account straightforwardly by extending the model and the state vector. As a next step, we will implement the mixed-control strategy in a test bench based on the real-time computer of the LBT-AO system. We should note that the mixed-control strategy can be implemented without changes on the existing hardware and firmware.

%\section{ACKNOWLEDGMENTS}

\bibliography{biblio}
\bibliographystyle{IEEEtranS}

\end{document}